\documentclass{article}

\usepackage[letterpaper,top=2cm,bottom=2cm,left=3cm,right=3cm,marginparwidth=1.75cm]{geometry}

\usepackage{amsmath}
\usepackage{graphicx}
\usepackage[colorlinks=true, allcolors=blue]{hyperref}

\usepackage{authblk}
\usepackage[numbers]{natbib}
\usepackage{amsmath,amssymb,amsfonts}
\usepackage{amsthm}
\usepackage{cite}
\usepackage{lmodern}

\providecommand{\keywords}[1]
{
  \small	
  \textbf{\textit{Keywords---}} #1
}

\title{Droplet impact on elastic substrates: force scaling crossover}
\author[1,2]{Yuto Yokoyama}
\author[3]{Hirokazu Maruoka}
\author[2,4]{Yoshiyuki Tagawa*}

\affil[1]{Micro/Bio/Nanofluidics Unit, Okinawa Institute of Science and Technology, 1919-1 Onna-son, Kunigami-gun, 9040497, Okinawa, Japan}

\affil[2]{Department of Mechanical Systems Engineering, Tokyo University of Agriculture and Technology, 2-24-16 Nakacho, Koganei, 1848588, Tokyo, Japan}

\affil[3]{Nonlinear and Non-equilibrium Physics Unit, Okinawa Institute of Science and Technology, 1919-1 Onna-son, Kunigami-gun, 9040497, Okinawa, Japan}

\affil[4]{Institute of Global Innovation Research, Tokyo University of Agriculture and Technology, 2-24-16 Nakacho, Koganei, 1848588, Tokyo, Japan}

\date{}

\begin{document}

\maketitle

\begin{abstract}
Droplet impacts are fundamental to fluid-structure interactions, shaping processes from erosion to bioprinting. While previous scaling laws have provided insights into droplet dynamics, force scaling laws remain insufficiently understood, particularly for soft substrates where both the droplet and substrate deform significantly. Here, we show that droplet impacts on elastic substrates exhibit a scaling crossover in maximum impact force, transitioning from inertial force scaling, typical for rigid substrates under high inertia, to Hertzian impact scaling, characteristic of rigid spheres on elastic substrates. Using high-speed photoelastic tomography, we captured high-resolution dynamic stress fields and identified a similarity parameter governing the interplay between droplet inertia, substrate elasticity, and deformation time scales. Our findings redefine how substrate properties influence impact forces, demonstrating that droplets under high inertia—long thought to follow inertial force scaling—can instead follow Hertzian impact scaling on soft substrates. This framework provides practical insights for designing soft, impact-resistant materials.
\end{abstract}

\keywords{Droplet Impact, Impact Force, Impact Stress, Substrate Elasticity, Scaling}
\vspace{8mm}

The behaviour of droplets impacting solid surfaces affects the quality and efficiency of many industrial processes, such as inkjet printing \citep{lohse2022} and spray cooling \citep{breitenbach2018,liang2017}.
The ubiquity and importance of droplet impact make it one of the most extensive areas of research in the fluid mechanics field \citep{josserand2016,yarin2006,yarin2017} over a period of about 150 years, starting since Worthington's pioneering work \citep{worthington1877,worthington1877a}.
While research has frequently concentrated on the morphology of droplets at the moment of impact \citep{liu2021a,liu2014}, it has been demonstrated that scaling laws are crucial in explaining these complex phenomena.
Scaling laws for droplet contact time \citep{richard2002,bird2013} and maximum spreading diameter \citep{clanet2004a,laan2014} have provided valuable insights into droplet dynamics.
Despite the importance of forces and stresses generated during droplet collisions, their dynamics remain poorly understood.
The impact force and stress produced by droplet impact are essential in various industrial technologies, such as aircraft erosion by raindrops \citep{hoksbergen2023}, cutting techniques such as water cutters \citep{mitchell2024}, and next-generation printing, such as bioprinting \citep{matai2020}.
In particular, the maximum impact force is an essential parameter for many impact-related processes, and its quantitative prediction is required. There has been extensive research on the factors determining the maximum impact force $F_{\rm max}$ \citep{zhang2022b, zhang2017, mitchell2019a}.
\citet{soto2014} experimentally demonstrated that $F_{\rm max}$ is proportional to the inertial force of the droplet in the high-Reynolds-number regime ($Re= \rho V R/\eta$, where $\rho$, $\eta$, $V$, and $R$ are the density, viscosity, impact velocity, and initial radius of the impacting droplet, respectively), i.e.
\begin{equation}\label{eq:inertial_force_scaling}
F_{\rm max} \propto \rho V^2 R^2.
\end{equation}
This indicates that the dimensionless maximum impact force, $\Tilde{F}_{\rm max}=F_{\rm max}/(\rho V^2 R^2)$, is constant in the high-$Re$ limit, which is called \textit{inertial force scaling} in the review by \citet{cheng2022}.
On the other hand, when the viscosity of the droplet increases, the fluid collides with the substrate with smaller spreads, but the normal force is significant, i.e. $F_{\rm max}$ increases with decreasing droplet viscosity in the low-$Re$ regime \citep{zhang2017}.
The variation of $F_{\rm max}$ with respect to $Re$ was formulated by \citet{gordillo2018}.
However, the data for $F_{\rm max}$ deviate from their predictions at $Re \lesssim 1$, and they mentioned that $Re$ is no longer a suitable dimensionless number for scaling $\tilde{F}_{\rm max}$ in this region.
Additionally, if the substrate is elastic, it can be supposed that the substrate deformation also becomes significant as viscosity increases.
Therefore, it is suggested that the substrate elasticity should be considered when the droplet viscosity is significantly high.
In this situation, $F_{\rm max}$ is expected to exhibit \textit{Hertzian impact scaling}, i.e.
\begin{equation}\label{eq:hertzian_impact_scaling}
F_{\rm max}\propto \rho^{3/5} V^{6/5} R^2 E^{2/5},
\end{equation}
as observed when a rigid sphere impacts an elastic substrate with an elastic modulus of $E$ \citep{johnson1985,yokoyama2024}.
However, the transition process of scaling law has never been observed, and the effect of the substrate elasticity on the maximum impact force is not clarified.

Investigating the stress field in the substrate, which changes in spatial distribution depending on the impacting droplet behaviour, is also crucial for understanding such a phenomenon involving fluid-structure interaction.
Nonetheless, experimental measurement of the stress field in the substrate during droplet impact was not achieved until very recently due to the requirement for high spatio-temporal resolution.
\citet{sun2022a} were the first to successfully measure the stress field in the elastic substrate during droplet impact using digital image correlation.
However, their experimental conditions were limited, and the effect of the substrate elasticity on the maximum impact force remains unclear.
Their contribution has enabled quantitative investigation of the stress field, opening up new possibilities for research on such a dynamic fluid-structure interaction problem.
However, to understand the impact force and stress in a wider parameter space, it is necessary to develop new stress measurement methods in addition to the above technique, as indicated in the latest review \citep{cheng2022}.

Here, we applied our novel optical stress measurement technology called ``high-speed photoelastic tomography'' \citep{yokoyama2024} to quantify the stress field in the elastic substrate during droplet impact. 
We clarified for the first time that as the droplet’s viscosity increases, the scaling law of the maximum impact force $F_{\rm max}$ shows a crossover from inertial force scaling to Hertzian impact scaling based on the stress field visualization.
To explain the physics behind this crossover and how the substrate elasticity $E$ is involved in the process, we explored the combination of similarity parameters bridging two asymptotics by the data-driven algorithm.

\begin{figure}
\centering
\includegraphics[width=1\columnwidth]{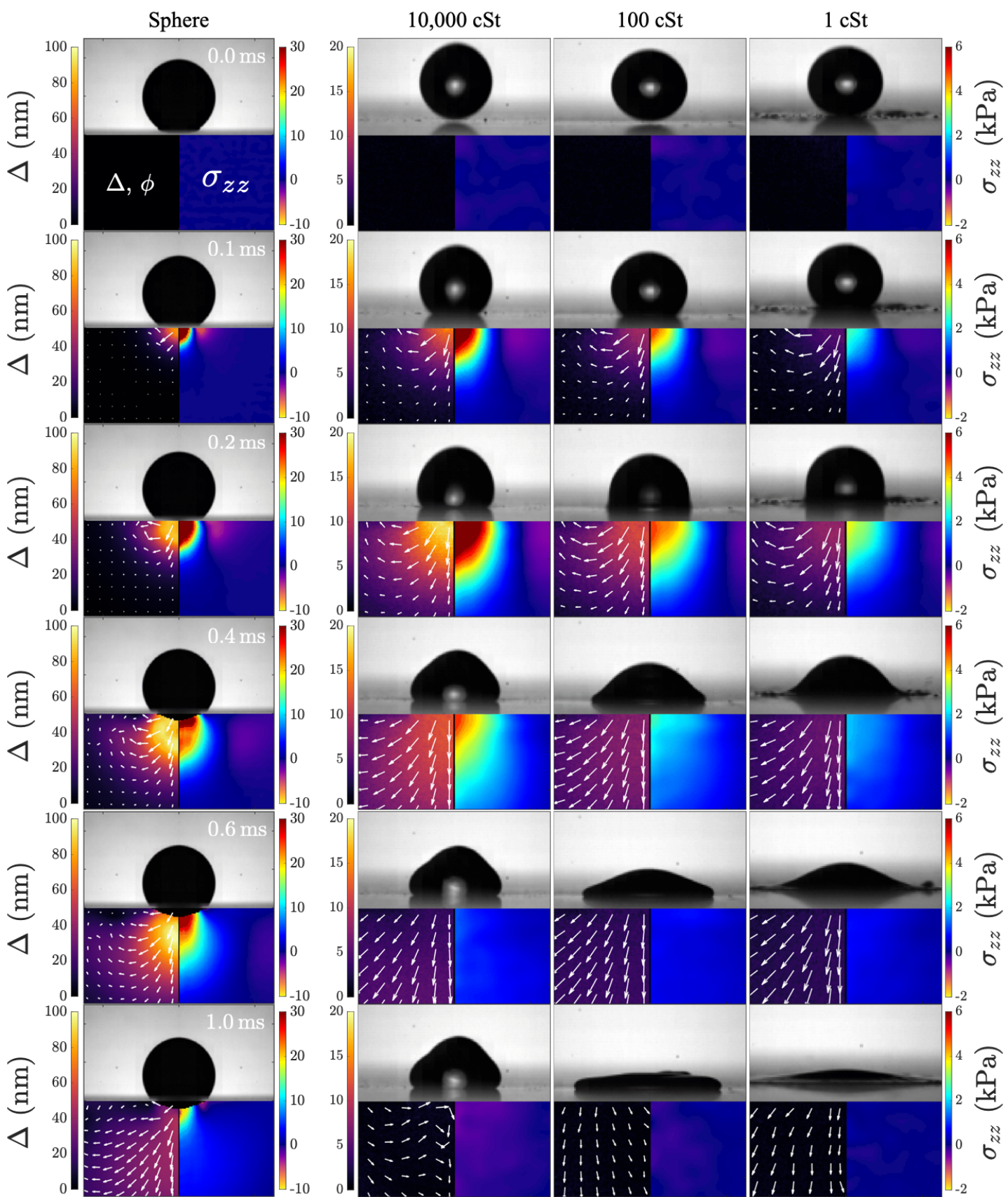}
\caption{\textbf{Dynamic behaviours of the impactors and the substrate response.} Spatio-temporal distribution of the stress field (bottom-left panels) photoelastic parameters, $\Delta$ and $\phi$, and (bottom-right panels) reconstructed axial stress $\sigma_{zz}$ when a sphere and droplet impact on an elastic substrate with $V \simeq 2.8 \pm 0.1$ m/s. In the bottom-left panels, the colourmap indicates the retardation $\Delta$ and white arrows indicate the orientation $\phi$.}
\label{fig:Sphere_and_Droplet_Impact_O_zz}
\end{figure}

\begin{figure}
\centering
\includegraphics[width=1\columnwidth]{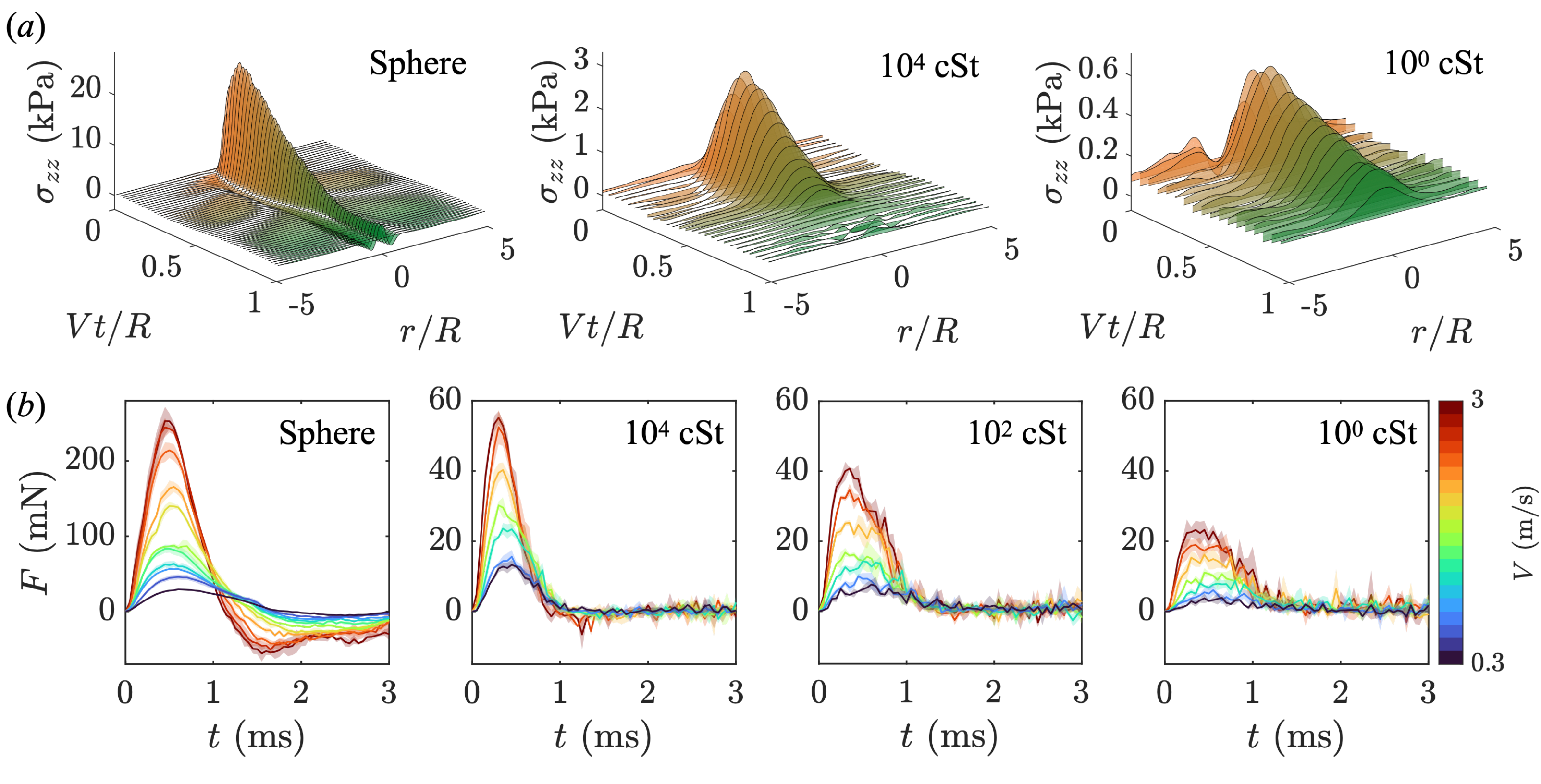}
\caption{\textbf{Stress and force acting on the elastic substrate.} (\textit{a}) Temporal evolution of the axial stress distribution acting on the substrate surface when Cauchy number, $Ca = \rho V^2/E$, is approximately 0.02. (\textit{b}) Temporal evolution of the impact force with different impact velocities during a sphere or droplet impact. The shaded regions represent one standard deviation for three experiments.}
\label{fig:ImpactForce}
\end{figure}

In the experiment, we recorded the behaviour of the impactor (a rigid sphere or silicone oil droplets with different viscosity) and the transparent elastic substrate (a polyurethane gel) by using the high-speed polarization camera.
The polarization camera can access the polarization state, which is called photoelastic parameters, of the light passing through the stressed substrate.
The polarization state is used for the stress reconstruction based on photoelastic tomography (see Method for details).
Figure \ref{fig:Sphere_and_Droplet_Impact_O_zz} shows the sequence of greyscale images of the impactor (top), the photoelastic parameters ($\Delta$ and $\phi$, bottom left), and the reconstructed axial stress ($\sigma_{zz}$, bottom right).
The time at which the impactor touches the substrate is defined as $t = 0 \, \rm s$.
Greyscale images show that, after impact, the highly viscous droplets do not spread much due to the viscous dissipation \citep{pasandideh-fard1996}, whereas the low-viscosity droplets spread significantly.
When impacted by the rigid sphere, the substrate deforms up to approximately 40\% of the sphere radius.
The substrate deformation during droplet impact is so slight that it cannot be clearly observed with the current spatial resolution of the image.

The retardation $\Delta$, which corresponds to the integration of secondary principal stress difference within the substrate, is larger for sphere impact than for droplet impact, as shown by the difference in the colour bar in figure \ref{fig:Sphere_and_Droplet_Impact_O_zz}.
This means that the stress induced by the sphere impact is much larger than that of the droplet impact.
In the case of droplets, the maximum retardation, i.e. maximum stress, increases with the droplet viscosity.
The orientation $\phi$ (white arrows), which is related to the direction of secondary principal stress within the substrate, is directed downwards overall in the area below the droplet-substrate contact.
In response to this, it is directed upwards in the outer region of the area.
This behaviour means that the substrate is deforming as an elastic-half space rather than as a simple Winkler elastic foundation, which does not have the interaction between the adjacent elements in a horizontal direction \citep{johnson1985}.

In the bottom-left panels of figure \ref{fig:Sphere_and_Droplet_Impact_O_zz}, along the $r$-axis of $\sigma_{zz}$ at $t \simeq 0.1$ ms, in the region near the contact area, $\sigma_{zz}$ is positive, whereas in the outer region, $\sigma_{zz}$ is negative.
Positive values of $\sigma_{zz}$ indicate that the substrate is compressed in the $z$-axis, and negative values indicate tensile deformation in the $z$-axis.
Immediately after impact, the boundary between positive and negative stresses follows the rim of the contact area.
It then propagates in the $r$-direction beyond the contact area.
The maximum stress of $\sigma_{zz}$ appears approximately 0.2 ms after impact, and its value increases with increasing droplet viscosity.
At $t = 1.0 \, \rm ms$, the highly viscous droplets, e.g., $10^4$ cSt silicone oil, produce negative values of $\sigma_{zz}$ even below the droplet--substrate contact area.
This is because the substrate, which has been slightly deformed downwards by droplet impact, deforms upwards to return to its original position due to its elasticity.
This negative stress is more pronounced in the sphere case and is barely noticeable in low-viscosity droplets, which produce little deformation of the substrate.

Figure \ref{fig:ImpactForce}(\textit{a}) shows the change in axial stress distribution on the substrate surface over time. 
In the case of the sphere, the stress region near the centre remains narrow because the sphere cannot spread.
The $10^4$ cSt silicone oil droplet exhibits a similar behaviour due to its high viscosity, which prevents rapid spreading along the $r$-axis.
In contrast, the $10^0$ cSt silicone oil droplet shows a wider propagation of stress distribution along the $r$-axis because of the faster spreading of the droplet.
However, it does not show the non-central stress peak, which is predicted by many theoretical and numerical works \citep{mandre2009,howland2016,philippi2016,hoksbergen2023,fu2024} and experimentally observed by \citet{sun2022a}.
\citet{sun2022a} indicated that if the substrate is elastic, the peak position of $\sigma_{zz}$ becomes closer to the centre, $r = 0$, which is not expected that the substrate is infinitely rigid.
Therefore, since the substrate we use is much softer than that used in their work, the non-central stress peak may merge with the stressed region around the centre.

As shown in figure \ref{fig:ImpactForce}(\textit{b}), for each droplet impact, the maximum value of the impact force, $F_{\rm max}$, increases with increasing impact velocity $V$, i.e. droplet inertia, and droplet viscosity $\eta$.
For the sphere case, $F(t)$ is symmetric with respect to the time when $F_{\rm max}$ appears if $V$ is sufficiently high.
This is expected under Hertzian impact theory \citep{johnson1985}.
$F(t)$ is also symmetric for the high-viscosity droplets.
However, for low-viscosity droplets, $F(t)$ exhibits an asymmetric shape.
These tendencies are similar to the results of previous studies \citep{zhang2017,gordillo2018}, although this earlier work used non-deformable rigid substrates, whereas the present experiment uses a deformable elastic substrate.

Next, we discuss the scaling of our problem, i.e., how the maximum impact force is influenced by the physical quantities, especially the droplet viscosity and the substrate elasticity.
The physical parameters of interest are $F_\mathrm{max}, \rho, V, R, \eta$, and $E$.
Let $F_\mathrm{max}$ be described as a function, $F_{\rm max}=g\left( \rho, V, R, \eta,E \right)$.
We introduce the following nondimensionalization.
By selecting $\rho, E, R$ as physical parameters with independent dimensions \citep{barenblatt2003}, we naturally obtain the relationship between the dimensionless numbers, $\Pi = f(\theta,Ca)$, where $\Pi= F_\mathrm{max}/ (E R^2)$, $Ca = \rho V^2 / E$, and $\theta = \eta/ \sqrt{\rho E R^2}$.
This combination of dimensionless numbers effectively separates $V$ and $\eta$ \citep{maruoka2023}, which are varied in the experiment.
The relationships between them and the dimensionless numbers $\tilde{F}_{\rm max}$ and $Re$ are $\Pi = \tilde{F}_{\rm max} Ca$ and $\theta = \sqrt{Ca}/Re$.

\begin{figure}
\centering
\includegraphics[width=1\columnwidth]{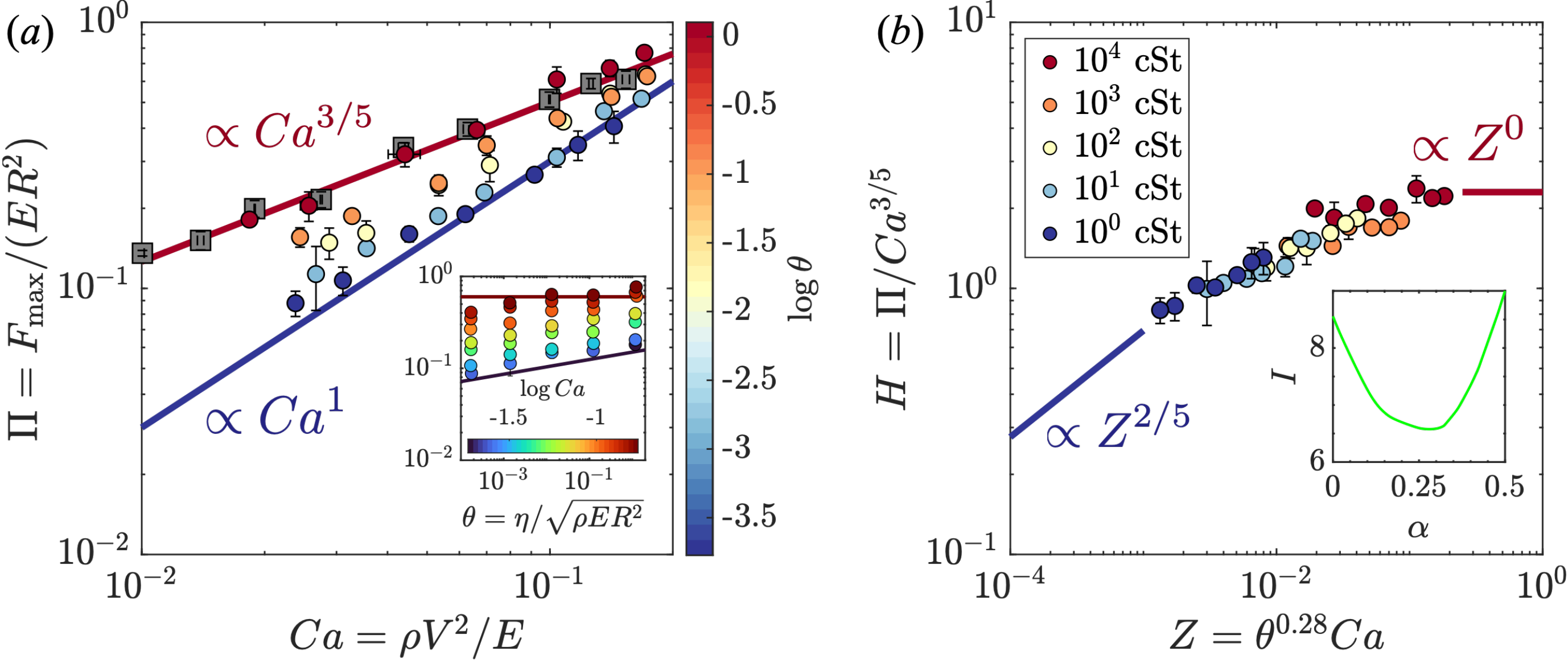}
\caption{\textbf{Self-similar structure of the maximum impact force.} (\textit{a}) $\Pi$ vs. $Ca$. The grey square markers represent the data of the sphere case. The inset of (\textit{a}) represents $\Pi$ vs. $\theta$. (\textit{b}) $H$ vs. $Z = \theta^\alpha Ca$ with $\alpha = 0.28$. The red solid line indicates Hertzian impact scaling of the form of (\ref{eq:HertzSim}) and the blue solid line indicates inertial force scaling of the form of (\ref{eq:DropSim}). The inset of (\textit{b}) shows $I$ vs. $\alpha$. The error bars represent one standard deviation over three impacts.}
\label{fig:SelfSimi}
\end{figure}

Figure \ref{fig:SelfSimi}($a$) shows the scaling relations between $\Pi$ and $Ca$, illustrating the different power-law behaviours depending on $\theta$, and two distinct behaviours can be found.
The first is Hertzian impact scaling \citep{johnson1985}, where $\Pi \propto Ca^{3/5}$ for large $\theta$; see also Eq. (\ref{eq:hertzian_impact_scaling}).
The second is inertial force scaling \citep{cheng2022}, where $\Pi \propto Ca$ for small $\theta$; see also Eq. (\ref{eq:inertial_force_scaling}).
This interpretation is consistent pictures with the observation on the stress and force shown in figures \ref{fig:Sphere_and_Droplet_Impact_O_zz} and \ref{fig:ImpactForce}.
Furthermore, as the plots do not collapse under these dimensionless numbers, the problem is likely to fall into the self-similarity of the second kind, for which the self-similarity cannot be captured solely through dimensional analysis (see Method for the detailed explanation about the self-similarity).

As the scaling behaviour changes depending on $\theta$, including $\eta$, $\theta$ is a driving parameter that governs the crossover of scaling law.
In the limit of larger $\theta$ (or larger $Ca$), the scaling follows the Hertzian impact scaling.
Therefore, the function, $f$, must satisfy the following behaviour in this limit, $f(\theta,Ca)  \propto Ca^{3/5}.$
To consider the self-similar solution that governs the crossover, we define the following dimensionless number and the function, $H = \Pi/Ca^{3/5}$ and $\Phi\left(\theta, Ca \right)= f\left(\theta, Ca \right) / Ca^{3/5}$, according to the framework of \citet{maruoka2023}.
In the Hertzian impact scaling as large $\theta$, $\Phi$ must satisfy
\begin{equation}\label{eq:HertzSim}
\Phi \left( \theta, Ca \right) \propto {\rm const.}~\left( \theta \gg 1 \right).
\end{equation}
On the other hand, the inertia force scaling ($\Pi \propto Ca$) is recovered in small $\theta$.
Thus $\Phi$ follows
\begin{equation}\label{eq:DropSim}
\Phi \left( \theta, Ca \right) \propto  Ca^{2/5}~\left( \theta \ll 1 \right).
\end{equation}
$\Phi$ should satisfy these two asymptotic behaviours while $\Phi$ is still indeterminate.
However, assuming that $\Phi$ belongs to the similarity of the second kind, $\Phi$ has the potential to have the following form: $\Phi\left(\theta, Ca \right)  = \Phi\left(\theta^\alpha Ca \right)$.
Using the prayer beads algorithm (see Method section) in which the power exponent of the similarity parameter, $\alpha$, was optimized so that the sum of the distance of each data plots $I\left(\alpha \right)$ is minimized, $\alpha$ was estimated as $\alpha \simeq 0.28$ while the optimized values should be located around 0.15--0.35, as shown in the inset of figure \ref{fig:SelfSimi}(\textit{b}). 
Plotting the data points using the similarity parameters $H$ and $Z = \theta^\alpha Ca$ as figure \ref{fig:SelfSimi}(\textit{b}), the plot reveals seemingly good data collapse, and the asymptotic behaviour follows the condition for the crossover of scaling laws, i.e. (\ref{eq:HertzSim}) and (\ref{eq:DropSim}).

At present, the theoretical interpretation of $\alpha$ is an open question.
However, we can consider the physical meaning of the optimized dimensionless number $Z = \theta^{\alpha}Ca$, which provides the data collapse for $H$.
$Z$ can be decomposed as follows:
$ \theta^{\alpha}Ca = \left( \frac{\eta/E}{R/V} \right)^{\alpha} \left(\frac{\rho V^2}{E}\right)^{1-\frac{1}{2}\alpha}
= \rm \left( \frac{Relaxation \, time}{Contact \, time} \right)^{\alpha} \left(\frac{Inertial \, force}{Elastic \, force} \right)^{1-\frac{1}{2}\alpha}$.
Here, we consider $\eta/E$ as the relaxation time associated with the droplet and substrate surface deformation and $R/V$ as the time duration in which a droplet spreads on the substrate, namely the contact time.
The time scale ratio can also be expressed using the Deborah number.
Therefore, this combination of dimensionless numbers reflects the ratio between the contact time of the droplet and the relaxation time of droplet--substrate deformation, as well as the ratio of substrate elastic force to droplet inertial force.
When an increase in droplet viscosity suppresses the spreading of the droplet, the relaxation time becomes significantly longer than the contact time, resulting in the scaling law approaching Hertzian impact scaling.
Conversely, when a decrease in droplet viscosity accelerates the droplet spreading, the relaxation time becomes shorter than the contact time, leading to a deviation from Hertzian impact scaling and resulting in inertial force scaling.
This is potentially considered a key mechanism for determining the maximum impact force of droplets on elastic substrates.
Note that we have defined the relaxation time as the ratio of droplet viscosity to substrate elastic modulus, $\eta/E$, although this quantity is normally expressed using the properties of the same material.
The reason for this choice is that the droplet and substrate are always in contact until the maximum impact force appears.

Based on our findings, earlier research has demonstrated that the scaling law for the maximum impact force with a rigid substrate is predominantly governed by $Re$, which reflects the balance between inertia and viscosity. 
This approaches $\Pi \propto Ca$ if the inertia is much greater than the viscosity and is named ``inertial force scaling'' \citep{cheng2022}.
In contrast, with an elastic substrate, increasing the inertia resulting in an increase in $Z$, shifts the scaling law from inertial force scaling to Hertzian impact scaling, even if the droplet viscosity is low.
This is because the scaling is influenced by the interplay between inertial and elastic forces and the comparison between relaxation time and contact time.
Thus, under this condition, i.e. with an elastic substrate, it would be misleading to continue referring to the scaling law $\Pi \propto Ca$ as ``inertial force scaling''.

In this paper, we successfully observed the transition of the dynamic behaviour of the impacting droplet depending on the droplet viscosity and the substrate elasticity through the measurement of stress and impact force acting on the elastic substrate using the photoelastic tomography technique.
We confirmed that this transition is from inertial force scaling to Hertzian impact scaling based on observing the stress distributions and the maximum impact force.
Our findings introduce a new insight for predicting the behaviour of droplets upon impact with substrates, revealing understandings crucial for both fundamental and applied science.
Based on our results—particularly figure \ref{fig:SelfSimi}(\textit{b})—we can possibly anticipate whether a droplet will behave like a rigid sphere or remain more fluidic upon impact based on parameters such as viscosity, inertia, and substrate elasticity.
This predictive capability has significant implications for applications where impact forces dictate material response. 
In turbine erosion or water-cutting, low-viscosity water droplets exhibit like solid spheres under high inertia, allowing them to erode or cut hard surfaces, including metals. 
This finding challenges conventional models, emphasizing the need to consider substrate elasticity to accurately predict the droplet impact force.
Furthermore, we believe that these insights have impacted the applications of soft materials. 
For instance, our ability to assess droplet impact behaviour enables more precise control in applications like 3D bioprinting, where impact stress affects the fidelity and viability of printed structures.
This study thus marks a step forward in comprehending fluid-structure interactions under high-speed impacts, bridging theoretical advances with practical application across diverse fields.

\section*{Method}\label{Sec:Method}

\subsection*{Experiment}

\begin{figure}
\centering
\includegraphics[width = 0.9\columnwidth]{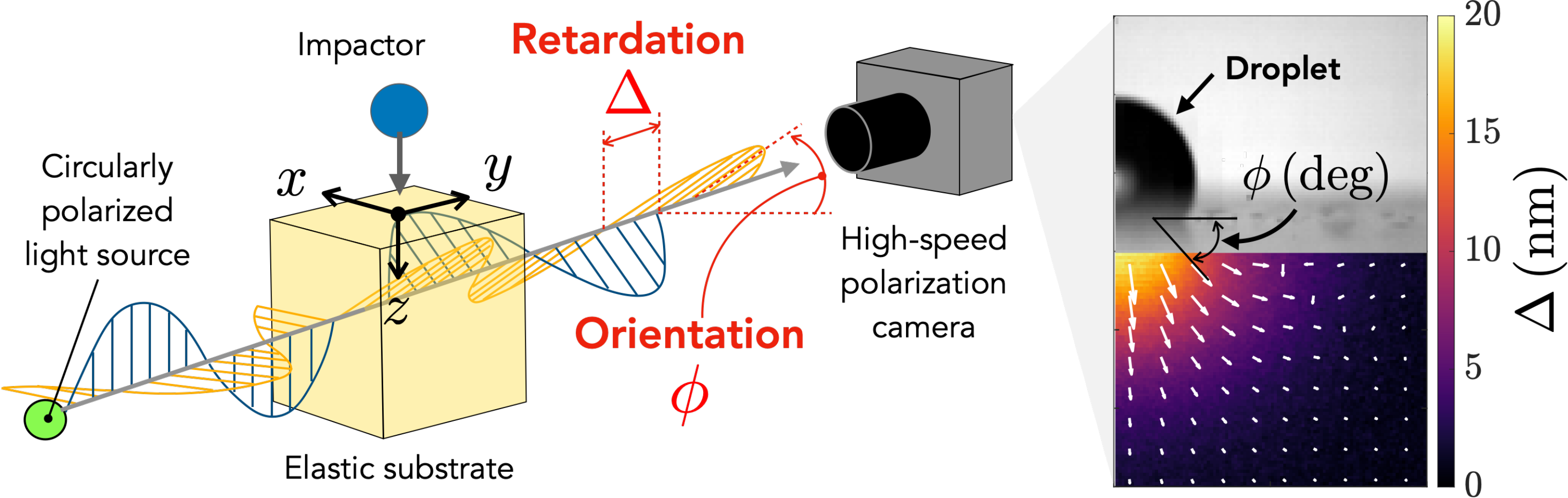}
\caption{\textbf{Stress field measurement using high-speed photoelastic tomography.} Schematic of the experimental setup for measuring the stress field in an elastic substrate during a sphere or droplet impact. The right panel shows a typical image of the photoelastic parameters (retardation $\Delta$ and orientation $\phi$) during the droplet impact. The colour indicates $\Delta$ and the white arrows indicate $\phi$.}
\label{fig:setup}
\end{figure}

The experimental setup is shown in figure \ref{fig:setup}.
Droplets of silicone oil (Shin-Etsu, KF-96) with kinetic viscosities of $10^0, 10^1, 10^2, 10^3, 10^4$ cSt were used.
The oil density $\rho$ was successively set to 815.5, 932.2, 962.1, 967.1, and 972.1 $\rm kg/m^3$.
The average droplet radius $R$ was $1.27$ mm.
The rigid sphere was made of plastic and had a radius of $R = 2.98$ mm and a density of $\rho = 1064$ $\rm kg/m^3$.
A gel block (Exseal Co., Ltd., polyurethane gel phantom, $50\times50\times50 \, \rm mm^3$) with an elastic modulus $E$ of 47.4 kPa was used as an elastic substrate.
The droplet or sphere impacted the substrate after falling freely.
The impact velocity $V$ was varied from approximately 0.3--3.0 m/s by adjusting the falling height from 1-50 cm.
To measure the impact force and stress within the substrate, photoelastic tomography was employed using the high-speed polarization camera (Photron, CRYSTA PI-1P, 20,000 fps) and the light source producing circularly polarized light (Thorlabs, SOLIOS-525C, typical wavelength of $\lambda = 520$ nm).
The measurements were repeated three times for each height.
The data for each height were averaged and plotted with their standard deviation in the following figures.

The photoelastic method measures the state of the polarized light passing through the stressed material.
This allows the stress field to be evaluated from the optical anisotropy (birefringence) caused by stress loading \citep{aben1993a}.
When circularly polarized light enters a stressed material, it is modulated as elliptically polarized light with photoelastic parameters (retardation $\Delta$ and orientation $\phi$) relating to the stress state.
$\Delta$ and $\phi$ can be measured by a polarization camera based on the four-step phase-shifting method \citep{onuma2014,yokoyama2023}.
Four linear micro-polarizers were installed in neighbouring pixels of the camera's image sensor.
The angles of the linear polarizers were set to $0^\circ$, $45^\circ$, $90^\circ$, and $135^\circ$, and the corresponding camera sensor measures the light intensity through these linear polarizers, denoted by $I_{0^\circ}, I_{45^\circ}, I_{90^\circ}$, and $I_{135^\circ}$, respectively.
$\Delta$ and $\phi$ can be obtained from the four intensity values as follows: $\Delta = \frac{\lambda}{2\pi} \sin^{-1}{\frac{\sqrt{\left(I_{90^\circ}-I_{0^\circ}\right)^2+\left(I_{45^\circ}-I_{135^\circ}\right)^2}}{(I_{0^\circ} + I_{45^\circ} + I_{90^\circ} + I_{135^\circ})/2}}, \phi = \frac{1}{2}\tan^{-1}\frac{I_{90^\circ}-I_{0^\circ}}{I_{45^\circ}-I_{135^\circ}}$.
A typical image of $\Delta$ and $\phi$ during droplet impact is shown on the right-hand side of figure \ref{fig:setup}.
The relationship between the stress field and photoelastic parameters is called the stress-optic law \citep{aben1993a}.
$\Delta$ is proportional to the integration of the secondary principal stress difference along the camera's optical axis, and $\phi$ is related to the direction of the secondary principal stress \citep{aben1993a}.
This relationship can be expressed as follows: $\Delta \cos 2 \phi = C \int^\infty_{-\infty} \left(\sigma_{xx}-\sigma_{zz}\right) dy, \Delta \sin 2 \phi = 2C \int^\infty_{-\infty} \sigma_{xz} dy,$
where $C$ is the stress-optic coefficient, which is $1.14\times 10^{-9}$ Pa$^{-1}$ for the substrate material used in this study \citep{yokoyama2024}.
The stress components in Cartesian coordinates are $\sigma_{xx}, \sigma_{zz}$, and $\sigma_{xz}$, with the $y$-axis as the camera's optical axis, as shown in figure \ref{fig:setup}.
From $\Delta$ and $\phi$, the dynamic stress field in the substrate can be reconstructed using our recently developed high-speed photoelastic tomography technique.
The impact force $F$ was estimated by integrating the reconstructed $\sigma_{zz}$ acting on the substrate's surface, i.e. $F(t) = 2 \pi \int^\infty_0 \sigma_{zz} r dr$.

\subsection*{Prayer beads algorithm for data collapse}\label{sec:data collapse algorithm}

To identify the self-similar solution to describe the crossover of scaling law \citep{maruoka2023,watanabe2024}, here we briefly review an algorithm to get data collapse.

Suppose that there are physical parameters $u, q, T$ and a function $u = f\left(q, T\right)$.
Here, we think about the operation to change the scale of parameters preserving the similarity; it is called scale-transformation, e.g., $(u', q', T') = (Au, A^\beta q, A^\gamma T)$ where $A$ is a factor and $\beta$, $\gamma$ are power exponents. We can deduce {\it invariant functions} of the scale-transformation, which is the form of the function such that does not vary by the scale-transformation.
It was known that invariant functions of scale-transformation are always power-law monomials \citep{barenblatt2003,olver1986}, e.g. $q/u^\beta, T/u^{\gamma}$.
Therefore, when the parameters of the function are replaced by the invariant form such as $q/u^{\beta} = \Phi\left(T/u^{\gamma} \right)$ where $\Phi$ is called a {\it scaling function}, and $q/u^{\beta}$ and $T/u^{\gamma}$ are called {\it similarity parameter}, the function $\Phi$ does not change by the scale-transformation as well.
It means that replotting the data points by using similarity parameters renormalizes the difference of the scale of the parameters to all the data points converge to a single line described by $\Phi$; it is {\it data collapse}, which is the expression exploiting the self-similarity and offering the fundamental insight about the problem.

This insight suggests that the condition of data collapse can be realized by identifying the power exponents of similarity parameters.
Dimensional analysis is a powerful method to get power exponents, which is called {\it similarity of the first kind}, though it is only applicable to special cases.
Because the problems belong to {\it self-similarity parameters of the second kind}, in which the dimensional analysis cannot deduce the power exponents. 
In such a case, the data collapse relation can be obtained by optimizing an appropriate evaluation function of the power exponents, $I\left(\beta, \gamma \right)$

Here, we propose a method to determine $\beta$ and $\gamma$ using the data-driven ``prayer beads algorithm''.
When $\beta$ and $\gamma$ are selected appropriately, all the scattering data points converge to a single curve described by the self-similar function $\Phi$.
At this time, the total distance between neighbouring data points is minimized.
Based on this observation, we define the evaluation function as the distance to each data point.
Here, either $\beta$ or $\gamma$ must be fixed to prevent data points from concentrating on a single point.
This picture corresponds to pulling the rope of relaxed prayer beads.
Supposing an exponent of one similarity parameter $\beta$ is known \emph{a priori} to be $\beta^\ast$, the evaluation function $I$ that optimizes $\gamma$ using $N$ data points can be written as follows:
\begin{equation}
I\left( \gamma \right)= \sum_i^{N} \sqrt{ (\log (q_{i+1}/u_{i+1}^{\beta^\ast}) - \log (q_{i}/u_{i}^{\beta^\ast}))^2 + (\log (T_{i+1}/u_{i+1}^{\gamma}) - \log (T_{i}/u_{i}^{\gamma}))^2 }.
\label{eq:DataCol}
\end{equation}
The optimized $\gamma^\ast$ is obtained when $I\left(\gamma^\ast \right) = \min I\left( \gamma\right)$.
In this study, the evaluation function of the problem was $I\left( \alpha \right) = \sum_i^{N}\sqrt{\left(\log (H_{i+1}/H_i)\right)^2 + \left(\log (Z_{i+1}/Z_i)\right)^2}$, where $H=\Pi/Ca^{3/5}$ and $Z = \theta^\alpha Ca$.



\section*{Acknowledgements}

This work was supported by JSPS KAKENHI Grant Numbers JP23KJ0859, JP24H00289, JP24KJ2176, and JST PRESTO Grant Number JPMJPR21O5, Japan. We thank Dr. D. Lohse, Dr. V. Sanjay, and Dr. S. Mandeep for their valuable discussions and suggestions. 





\bibliographystyle{unsrtnat}
\bibliography{ref}

\begin{thebibliography}{38}
\providecommand{\natexlab}[1]{#1}
\providecommand{\url}[1]{\texttt{#1}}
\expandafter\ifx\csname urlstyle\endcsname\relax
  \providecommand{\doi}[1]{doi: #1}\else
  \providecommand{\doi}{doi: \begingroup \urlstyle{rm}\Url}\fi

\bibitem[Lohse(2022)]{lohse2022}
Detlef Lohse.
\newblock Fundamental fluid dynamics challenges in inkjet printing.
\newblock \emph{Annual Review of Fluid Mechanics}, 54\penalty0 (1):\penalty0 349--382, 2022.
\newblock \doi{10.1146/annurev-fluid-022321-114001}.

\bibitem[Breitenbach et~al.(2018)Breitenbach, Roisman, and Tropea]{breitenbach2018}
Jan Breitenbach, Ilia~V. Roisman, and Cameron Tropea.
\newblock From drop impact physics to spray cooling models: A critical review.
\newblock \emph{Experiments in Fluids}, 59\penalty0 (3):\penalty0 55, February 2018.
\newblock ISSN 1432-1114.
\newblock \doi{10.1007/s00348-018-2514-3}.

\bibitem[Liang and Mudawar(2017)]{liang2017}
Gangtao Liang and Issam Mudawar.
\newblock Review of drop impact on heated walls.
\newblock \emph{International Journal of Heat and Mass Transfer}, 106:\penalty0 103--126, March 2017.
\newblock ISSN 0017-9310.
\newblock \doi{10.1016/j.ijheatmasstransfer.2016.10.031}.

\bibitem[Josserand and Thoroddsen(2016)]{josserand2016}
C.~Josserand and S.T. Thoroddsen.
\newblock Drop impact on a solid surface.
\newblock \emph{Annual Review of Fluid Mechanics}, 48\penalty0 (1):\penalty0 365--391, 2016.
\newblock \doi{10.1146/annurev-fluid-122414-034401}.

\bibitem[Yarin(2006)]{yarin2006}
A.L. Yarin.
\newblock Drop impact dynamics: Splashing, spreading, receding, bouncing{\dots}.
\newblock \emph{Annual Review of Fluid Mechanics}, 38\penalty0 (1):\penalty0 159--192, 2006.
\newblock \doi{10.1146/annurev.fluid.38.050304.092144}.

\bibitem[Yarin et~al.(2017)Yarin, Roisman, and Tropea]{yarin2017}
Alexander~L. Yarin, Ilia~V. Roisman, and Cameron Tropea.
\newblock \emph{Collision Phenomena in Liquids and Solids}.
\newblock Cambridge University Press, Cambridge, 2017.
\newblock ISBN 978-1-107-14790-4.
\newblock \doi{10.1017/9781316556580}.

\bibitem[Worthington and Stewart(1877)]{worthington1877}
Arthur~Mason Worthington and Balfour Stewart.
\newblock {{III}}. {{A}} second paper on the forms assumed by drops of liquids falling vertically on a horizontal plate.
\newblock \emph{Proceedings of the Royal Society of London}, 25\penalty0 (171-178):\penalty0 498--503, January 1877.
\newblock \doi{10.1098/rspl.1876.0073}.

\bibitem[Worthington and Clifton(1877)]{worthington1877a}
Arthur~Mason Worthington and Robert~Bellamy Clifton.
\newblock {{XXVIII}}. on the forms assumed by drops of liquids falling vertically on a horizontal plate.
\newblock \emph{Proceedings of the Royal Society of London}, 25\penalty0 (171-178):\penalty0 261--272, January 1877.
\newblock \doi{10.1098/rspl.1876.0048}.

\bibitem[Liu et~al.(2021)Liu, Lo, Li, Liu, Zhao, and Xu]{liu2021a}
Qingzhe Liu, Jack Hau~Yung Lo, Ye~Li, Yuan Liu, Jinyu Zhao, and Lei Xu.
\newblock The role of drop shape in impact and splash.
\newblock \emph{Nature Communications}, 12\penalty0 (1):\penalty0 3068, May 2021.
\newblock ISSN 2041-1723.
\newblock \doi{10.1038/s41467-021-23138-4}.

\bibitem[Liu et~al.(2014)Liu, Moevius, Xu, Qian, Yeomans, and Wang]{liu2014}
Yahua Liu, Lisa Moevius, Xinpeng Xu, Tiezheng Qian, Julia~M. Yeomans, and Zuankai Wang.
\newblock Pancake bouncing on superhydrophobic surfaces.
\newblock \emph{Nature Physics}, 10\penalty0 (7):\penalty0 515--519, July 2014.
\newblock ISSN 1745-2481.
\newblock \doi{10.1038/nphys2980}.

\bibitem[Richard et~al.(2002)Richard, Clanet, and Qu{\'e}r{\'e}]{richard2002}
Denis Richard, Christophe Clanet, and David Qu{\'e}r{\'e}.
\newblock Contact time of a bouncing drop.
\newblock \emph{Nature}, 417\penalty0 (6891):\penalty0 811, June 2002.
\newblock ISSN 1476-4687.
\newblock \doi{10.1038/417811a}.

\bibitem[Bird et~al.(2013)Bird, Dhiman, Kwon, and Varanasi]{bird2013}
James~C. Bird, Rajeev Dhiman, Hyuk-Min Kwon, and Kripa~K. Varanasi.
\newblock Reducing the contact time of a bouncing drop.
\newblock \emph{Nature}, 503\penalty0 (7476):\penalty0 385--388, November 2013.
\newblock ISSN 1476-4687.
\newblock \doi{10.1038/nature12740}.

\bibitem[Clanet et~al.(2004)Clanet, B{\'e}guin, Richard, and Qu{\'e}r{\'e}]{clanet2004a}
Christophe Clanet, C{\'e}dric B{\'e}guin, Denis Richard, and David Qu{\'e}r{\'e}.
\newblock Maximal deformation of an impacting drop.
\newblock \emph{Journal of Fluid Mechanics}, 517:\penalty0 199--208, September 2004.
\newblock ISSN 0022-1120, 1469-7645.
\newblock \doi{10.1017/S0022112004000904}.

\bibitem[Laan et~al.(2014)Laan, {de Bruin}, Bartolo, Josserand, and Bonn]{laan2014}
Nick Laan, Karla~G. {de Bruin}, Denis Bartolo, Christophe Josserand, and Daniel Bonn.
\newblock Maximum diameter of impacting liquid droplets.
\newblock \emph{Physical Review Applied}, 2\penalty0 (4):\penalty0 044018, October 2014.
\newblock \doi{10.1103/PhysRevApplied.2.044018}.

\bibitem[Hoksbergen et~al.(2023)Hoksbergen, Akkerman, and Baran]{hoksbergen2023}
T.~H. Hoksbergen, R.~Akkerman, and I.~Baran.
\newblock Liquid droplet impact pressure on (elastic) solids for prediction of rain erosion loads on wind turbine blades.
\newblock \emph{Journal of Wind Engineering and Industrial Aerodynamics}, 233:\penalty0 105319, February 2023.
\newblock ISSN 0167-6105.
\newblock \doi{10.1016/j.jweia.2023.105319}.

\bibitem[Mitchell et~al.(2024)Mitchell, Korkolis, and Kinsey]{mitchell2024}
Benjamin~R. Mitchell, Yannis~P. Korkolis, and Brad~L. Kinsey.
\newblock Erosion characteristics of water droplet machining.
\newblock \emph{Journal of Materials Processing Technology}, 327:\penalty0 118359, June 2024.
\newblock ISSN 0924-0136.
\newblock \doi{10.1016/j.jmatprotec.2024.118359}.

\bibitem[Matai et~al.(2020)Matai, Kaur, Seyedsalehi, McClinton, and Laurencin]{matai2020}
Ishita Matai, Gurvinder Kaur, Amir Seyedsalehi, Aneesah McClinton, and Cato~T. Laurencin.
\newblock Progress in {{3D}} bioprinting technology for tissue/organ regenerative engineering.
\newblock \emph{Biomaterials}, 226:\penalty0 119536, January 2020.
\newblock ISSN 0142-9612.
\newblock \doi{10.1016/j.biomaterials.2019.119536}.

\bibitem[Zhang et~al.(2022)Zhang, Sanjay, Shi, Zhao, Lv, Feng, and Lohse]{zhang2022b}
Bin Zhang, Vatsal Sanjay, Songlin Shi, Yinggang Zhao, Cunjing Lv, Xi-Qiao Feng, and Detlef Lohse.
\newblock Impact forces of water drops falling on superhydrophobic surfaces.
\newblock \emph{Physical Review Letters}, 129\penalty0 (10):\penalty0 104501, August 2022.
\newblock \doi{10.1103/PhysRevLett.129.104501}.

\bibitem[Zhang et~al.(2017)Zhang, Li, Guo, and Lv]{zhang2017}
Bin Zhang, Jingyin Li, Penghua Guo, and Qian Lv.
\newblock Experimental studies on the effect of reynolds and weber numbers on the impact forces of low-speed droplets colliding with a solid surface.
\newblock \emph{Experiments in Fluids}, 58\penalty0 (9):\penalty0 125, August 2017.
\newblock ISSN 1432-1114.
\newblock \doi{10.1007/s00348-017-2413-z}.

\bibitem[Mitchell et~al.(2019)Mitchell, Klewicki, Korkolis, and Kinsey]{mitchell2019a}
Benjamin~R. Mitchell, Joseph~C. Klewicki, Yannis~P. Korkolis, and Brad~L. Kinsey.
\newblock The transient force profile of low-speed droplet impact: Measurements and model.
\newblock \emph{Journal of Fluid Mechanics}, 867:\penalty0 300--322, May 2019.
\newblock ISSN 0022-1120, 1469-7645.
\newblock \doi{10.1017/jfm.2019.141}.

\bibitem[Soto et~al.(2014)Soto, Larivi{\`e}re, Boutillon, Clanet, and Qu{\'e}r{\'e}]{soto2014}
Dan Soto, Aur{\'e}lie Borel~De Larivi{\`e}re, Xavier Boutillon, Christophe Clanet, and David Qu{\'e}r{\'e}.
\newblock The force of impacting rain.
\newblock \emph{Soft Matter}, 10\penalty0 (27):\penalty0 4929--4934, June 2014.
\newblock ISSN 1744-6848.
\newblock \doi{10.1039/C4SM00513A}.

\bibitem[Cheng et~al.(2022)Cheng, Sun, and Gordillo]{cheng2022}
Xiang Cheng, Ting-Pi Sun, and Leonardo Gordillo.
\newblock Drop impact dynamics: Impact force and stress distributions.
\newblock \emph{Annual Review of Fluid Mechanics}, 54\penalty0 (1):\penalty0 null, 2022.
\newblock \doi{10.1146/annurev-fluid-030321-103941}.

\bibitem[Gordillo et~al.(2018)Gordillo, Sun, and Cheng]{gordillo2018}
Leonardo Gordillo, Ting-Pi Sun, and Xiang Cheng.
\newblock Dynamics of drop impact on solid surfaces: Evolution of impact force and self-similar spreading.
\newblock \emph{Journal of Fluid Mechanics}, 840:\penalty0 190--214, April 2018.
\newblock ISSN 0022-1120, 1469-7645.
\newblock \doi{10.1017/jfm.2017.901}.

\bibitem[Johnson(1985)]{johnson1985}
K.~L. Johnson.
\newblock \emph{Contact Mechanics}.
\newblock Cambridge University Press, Cambridge, 1985.
\newblock ISBN 978-0-521-34796-9.
\newblock \doi{10.1017/CBO9781139171731}.

\bibitem[Yokoyama et~al.(2024)Yokoyama, Ichihara, and Tagawa]{yokoyama2024}
Yuto Yokoyama, Sayaka Ichihara, and Yoshiyuki Tagawa.
\newblock High-speed photoelastic tomography for axisymmetric stress fields in a soft material: {{Temporal}} evolution of all stress components.
\newblock \emph{Optics and Lasers in Engineering}, 178:\penalty0 108224, July 2024.
\newblock ISSN 0143-8166.
\newblock \doi{10.1016/j.optlaseng.2024.108224}.

\bibitem[Sun et~al.(2022)Sun, {\'A}lvarez-Novoa, Andrade, Guti{\'e}rrez, Gordillo, and Cheng]{sun2022a}
Ting-Pi Sun, Franco {\'A}lvarez-Novoa, Klebbert Andrade, Pablo Guti{\'e}rrez, Leonardo Gordillo, and Xiang Cheng.
\newblock Stress distribution and surface shock wave of drop impact.
\newblock \emph{Nature Communications}, 13\penalty0 (1):\penalty0 1703, March 2022.
\newblock ISSN 2041-1723.
\newblock \doi{10.1038/s41467-022-29345-x}.

\bibitem[Pasandideh-Fard et~al.(1996)Pasandideh-Fard, Qiao, Chandra, and Mostaghimi]{pasandideh-fard1996}
M.~Pasandideh-Fard, Y.~M. Qiao, S.~Chandra, and J.~Mostaghimi.
\newblock Capillary effects during droplet impact on a solid surface.
\newblock \emph{Physics of Fluids}, 8\penalty0 (3):\penalty0 650--659, March 1996.
\newblock ISSN 1070-6631.
\newblock \doi{10.1063/1.868850}.

\bibitem[Mandre et~al.(2009)Mandre, Mani, and Brenner]{mandre2009}
Shreyas Mandre, Madhav Mani, and Michael~P. Brenner.
\newblock Precursors to splashing of liquid droplets on a solid surface.
\newblock \emph{Physical Review Letters}, 102\penalty0 (13):\penalty0 134502, March 2009.
\newblock \doi{10.1103/PhysRevLett.102.134502}.

\bibitem[Howland et~al.(2016)Howland, Antkowiak, {Castrej{\'o}n-Pita}, Howison, Oliver, Style, and {Castrej{\'o}n-Pita}]{howland2016}
Christopher~J. Howland, Arnaud Antkowiak, J.~Rafael {Castrej{\'o}n-Pita}, Sam~D. Howison, James~M. Oliver, Robert~W. Style, and Alfonso~A. {Castrej{\'o}n-Pita}.
\newblock It's harder to splash on soft solids.
\newblock \emph{Physical Review Letters}, 117\penalty0 (18):\penalty0 184502, October 2016.
\newblock \doi{10.1103/PhysRevLett.117.184502}.

\bibitem[Philippi et~al.(2016)Philippi, Lagr{\'e}e, and Antkowiak]{philippi2016}
Julien Philippi, Pierre-Yves Lagr{\'e}e, and Arnaud Antkowiak.
\newblock Drop impact on a solid surface: Short-time self-similarity.
\newblock \emph{Journal of Fluid Mechanics}, 795:\penalty0 96--135, May 2016.
\newblock ISSN 0022-1120, 1469-7645.
\newblock \doi{10.1017/jfm.2016.142}.

\bibitem[Fu et~al.(2024)Fu, Jin, Zhang, Xue, Guo, Yao, Gao, Wang, and Wen]{fu2024}
Zunru Fu, Haichuan Jin, Jun Zhang, Tianyou Xue, Qi~Guo, Guice Yao, Hui Gao, Zuankai Wang, and Dongsheng Wen.
\newblock Low-{{Pressure Pancake Bouncing}} on {{Superhydrophobic Surfaces}}.
\newblock \emph{Small}, 20\penalty0 (31):\penalty0 2310200, 2024.
\newblock ISSN 1613-6829.
\newblock \doi{10.1002/smll.202310200}.

\bibitem[Barenblatt(2003)]{barenblatt2003}
Grigory~Isaakovich Barenblatt.
\newblock \emph{Scaling}.
\newblock Cambridge {{Texts}} in {{Applied Mathematics}}. Cambridge University Press, Cambridge, 2003.
\newblock ISBN 978-0-521-82657-0.
\newblock \doi{10.1017/CBO9780511814921}.

\bibitem[Maruoka(2023)]{maruoka2023}
Hirokazu Maruoka.
\newblock A framework for crossover of scaling law as a self-similar solution: Dynamical impact of viscoelastic board.
\newblock \emph{The European Physical Journal E}, 46\penalty0 (5):\penalty0 35, May 2023.
\newblock ISSN 1292-895X.
\newblock \doi{10.1140/epje/s10189-023-00292-9}.

\bibitem[Aben and Guillemet(1993)]{aben1993a}
Hillar Aben and Claude Guillemet.
\newblock \emph{Photoelasticity of Glass}.
\newblock Springer Berlin Heidelberg, Berlin, Heidelberg, 1993.
\newblock ISBN 978-3-642-50073-2 978-3-642-50071-8.
\newblock \doi{10.1007/978-3-642-50071-8}.

\bibitem[Onuma and Otani(2014)]{onuma2014}
Takashi Onuma and Yukitoshi Otani.
\newblock A development of two-dimensional birefringence distribution measurement system with a sampling rate of 1.3mhz.
\newblock \emph{Optics Communications}, 315:\penalty0 69--73, March 2014.
\newblock ISSN 0030-4018.
\newblock \doi{10.1016/j.optcom.2013.10.086}.

\bibitem[Yokoyama et~al.(2023)Yokoyama, Mitchell, Nassiri, Kinsey, Korkolis, and Tagawa]{yokoyama2023}
Yuto Yokoyama, Benjamin~R. Mitchell, Ali Nassiri, Brad~L. Kinsey, Yannis~P. Korkolis, and Yoshiyuki Tagawa.
\newblock Integrated photoelasticity in a soft material: Phase retardation, azimuthal angle, and stress-optic coefficient.
\newblock \emph{Optics and Lasers in Engineering}, 161:\penalty0 107335, February 2023.
\newblock ISSN 0143-8166.
\newblock \doi{10.1016/j.optlaseng.2022.107335}.

\bibitem[Watanabe et~al.(2024)Watanabe, Ishii, Hirono, and Maruoka]{watanabe2024}
Ryota Watanabe, Takanori Ishii, Yuji Hirono, and Hirokazu Maruoka.
\newblock Data-driven discovery of self-similarity using neural networks, June 2024.

\bibitem[Olver(1986)]{olver1986}
Peter~J. Olver.
\newblock \emph{Applications of {{Lie Groups}} to {{Differential Equations}}}, volume 107 of \emph{Graduate {{Texts}} in {{Mathematics}}}.
\newblock Springer, New York, NY, 1986.
\newblock ISBN 978-1-4684-0276-6 978-1-4684-0274-2.
\newblock \doi{10.1007/978-1-4684-0274-2}.

\end{thebibliography}

\end{document}